\documentclass[11pt,a4paper]{article}

\usepackage[latin1]{inputenc}
\usepackage{amsmath}\DeclareMathOperator{\sech}{sech}
\usepackage{amsfonts,amssymb,graphicx,amsthm,listings,float}
\usepackage{caption,subcaption,color,chngcntr,cite,authblk,bbold,braket}
\usepackage[titletoc]{appendix}  
\usepackage[english]{babel} 
\usepackage[left=2.00cm, right=2.00cm, top=2.00cm, bottom=2.00cm]{geometry}

\usepackage{soul}
\usepackage{xcolor}

\title{Complex Supersymmetry in Graphene}

\author{Miguel Castillo-Celeita\footnote{mfcastillo@fis.cinvestav.mx} }
\author{Alonso Contreras-Astorga\footnote{alonso.contreras@conacyt.mx}}
\author{David J. Fern\'andez C.\footnote{david@fis.cinvestav.mx}}

\affil{\small $~^{* \ddag}$ Physics Department, Cinvestav, P.O. Box. 14-740, 07000 Mexico City, Mexico \\ 
$~^\dag$ CONACyT-Physics Department, Cinvestav, P.O. Box. 14-740, 07000 Mexico City, Mexico}

\date{}
\begin{document}
\maketitle

\begin{abstract}
This work analyzes monolayer graphene in external electromagnetic fields, which is described by the Dirac equation with minimal coupling. Supersymmetric quantum mechanics allows building new Dirac equations with modified magnetic fields. Here, we will use complex factorization energies and iterate the method in order to arrive at Hermitian graphene Hamiltonians. Finally, we compare these results with the matrix supersymmetric quantum mechanics approach.        
\end{abstract}

\section{Introduction}

In solid-state physics, graphene is the most recent discovery of allotropes of carbon. Despite its recent finding in lab, its band structure is known since 1947 \cite{w47}. When it is studied at low energies, the dispersion relation is linear in the momentum, thus the corresponding  Hamiltonian becomes the one for a massless Dirac electron. For years this material was considered unrealistic, because previous theoretical works suggested the instability of 2D crystals at finite temperature \cite{p35,l37,fv16}, but its existence was confirmed in 2004 \cite{Novoselov04a,Novoselov05}. Since then, there have been a huge amount of research on the topic from both, theoretical and experimental viewpoints. Moreover, since charge carriers in graphene can be modeled by the Dirac equation, it  has allowed us to explore with tabletop experiments interesting phenomena of $(2+1)$-dimensional quantum electrodynamics, as the Klein tunneling or the quantum Hall effect \cite{Zhang05,Katsnelson06,CastroNeto09,be08}. 

To confine or control the charge carriers in a graphene sample, electromagnetic fields have to be applied \cite{Lukose07,Stander09,Hartmann10,Kraft20}; it has been shown that mechanical deformations can be used as well for that purpose~\cite{Vozmediano10,nbo17,pdrv20,Betancur18,Contreras20a}. To explore and enrich the configurations where the Dirac equation can be solved exactly (or quasiexactly), supersymmetric quantum mechanics have been used from different approaches \cite{knn09,mf14,Jakubsky11,scro14,cosc14,cefe20,Junker20,nipesa03,Ioffe19,Correa17,fgo20,fgo21,Contreras20}. From a mathematical perspective, a charge carrier in a graphene layer under an external magnetic field obeys simply a Dirac equation of the form $H \Phi = E \Phi$, where $H$ is a $2 \times 2$ matrix Hamiltonian and $\Phi$ is a two-entry spinor. By writing the eigenvalue equation in components and decoupling the resulting system of differential equations, it is found that each component must fulfill a Schr\"odinger equation, where the potentials are supersymmetric partners from each other~\cite{knn09}. Supersymmetry allows, in general, to modify the spectra of the Dirac Hamiltonians through a set of parameters $\epsilon_i$, called factorization energies, that are closely related with the spectral modifications. The case where $\epsilon_i$ are real has been widely studied in the above cited references. In this work, we will focus in the case where the $\epsilon_i$ are complex parameters, but the Hermiticity of the Dirac Hamiltonians produced by the supersymmetric technique is not completely lost, thus producing Hamiltonians having real energy spectra~\cite{mhc13,mhe13}. We will study two different approaches, the first one where we take advantage of the Schr\"odinger equations satisfied by each component of the eigenspinor, and the second approach where the matrix nature of the Hamiltonian is used and exploited. 
This article is organized as follows: first we present details of the complex supersymmetric quantum mechanics applied to the Schr\"odinger equation as an introduction~(see Section~\ref{sec:susy}); in Section \ref{sec:Complex SUSY} we give a general framework to use the complex supersymmetry for the Dirac equation, and we illustrate this procedure with an interesting example; Section \ref{sec:Complex SUSY Matix} introduces a matrix approach to supersymmetry, which is a more general framework containing the treatment of Section \ref{sec:Complex SUSY}, and it is also illustrated with an example; finally, we summarize our conclusions in the last section.     

\section{Complex supersymmetric quantum mechanics}\label{sec:susy}

Supersymmetric Quantum Mechanics (SUSY) is a technique that allows us to find solutions of a new Schr\"odinger equation given that we know the solution of an initial Schr\"odinger equation~\cite{cks95,ba00,fe10,fefe05,junker19,fe19}. The simplest SUSY transformation involves a first-order differential operator that intertwines two Schr\"odinger Hamiltonians. The technique can add a new energy level below the ground state, delete the initial ground state, or produce isospectral Hamiltonians. Those modifications of the spectrum depend on the value of so called factorization energy $\epsilon$ and the election of an auxiliary function $u(x)$. Note that the value of $\epsilon$ becomes the position of the ground state in the generated Hamiltonian. On the other hand, through a second-order differential intertwining operator it is possible to build isospectral Hermitian Hamiltonians using a complex factorization energy $\epsilon$. Although, the goal of this work is to study the Dirac equation, for the sake of completeness let us give a brief review of the complex case of SUSY; more details can be found in \cite{fefe05}.

\subsection{First-order complex supersymmetry} \label{sec: 1SUSY}
 We start by considering the Schr\"odinger equation for a Hermitian Hamiltonian $h_0$  with a time-independent potential $v_0$: 
\begin{eqnarray} \label{Sec2 Schrodinger}
h_0 \phi =  \varepsilon \phi,  \quad\quad h_0= - \frac{d^2}{dx^2} + v_0(x). 
\end{eqnarray}
As a first step, we propose the intertwining relation~\cite{cks95,ba00,fefe05,fe10}
\begin{eqnarray}
h_1 l_1^+ = l_1^+ h_0,
\label{eq7}  
\end{eqnarray}
where 
\begin{eqnarray}
h_1= - \frac{d^2}{dx^2} + v_1(x), \qquad l_1^+ = - \frac{d}{dx} + \frac{u_1'}{u_1}, \label{eq8}
\end{eqnarray}
with $u_1=u_1(x)$ being an arbitrary function called seed solution or transformation function. By substituting Eq.~\eqref{eq8} into the intertwining relation \eqref{eq7} we find that $u_1$ and  $v_1$  must fulfill 
\begin{eqnarray} \label{eq9}
-u_1''+v_0 u_1 = \epsilon_1 u_1, \quad v_1(x)= v_0(x) - 2 \frac{d^2}{dx^2} \ln u_1, 
\end{eqnarray}
where $\epsilon_1$ is an integration constant called factorization energy. Notice that $u_1$ satisfies the initial Schr\"odinger equation for the factorization energy $\epsilon_1$, but we will not impose any physical boundary condition on this solution. It is important to remark also that if $h_0$ is Hermitian $v_0$ must be a real function, thus a natural assumption is that both $u_1$ and $\epsilon_1$ should be real. This will not be the case in this work, and we will refer to this case as \emph{complex supersymmetry}.

By applying the operator $l_1^+$ onto the solutions $\phi$ of Eq.~\eqref{Sec2 Schrodinger} 
we obtain solutions $\hat{\phi} \propto l_1^+ \phi$ of the new Schr\"odinger equation $h_1 \hat{\phi}=\varepsilon \hat{\phi}$. This is guaranteed by the intertwining relation \eqref{eq7}: the operator $l_1^+$ maps the space of solutions of $h_0 \phi = \varepsilon \phi$ onto the space of solutions of $h_1 \hat{\phi} = \varepsilon \hat{\phi}$. In general, the mapped functions $\hat{\phi}$ do not satisfy the boundary conditions and the potential $v_1$ is not regular. These conditions are strongly related with the election of the transformation function $u_1$. In first-order SUSY, $u_1$ must be nodeless in the domain of $v_0$ in order to produce a regular potential $v_1$. 

There is an intertwining relation making the opposite map, $h_0 l_1 = l_1 h_1$, where $l_1 = d/dx+u_1'/u_1$. If $u_1$ is a real function, this intertwining relation arises by taking formally the adjoint of Eq.~\eqref{eq7}. In complex supersymmetry, this relation remains valid, although for this case $l_1 \neq (l_1^+)^\dagger$. The operators $l_1$ and $l_1^+$ factorize the Hamiltonians $h_0$ and $h_1$ as follows:
\begin{eqnarray} \label{factorization1}
l_1 l_1^+ = h_0 - \epsilon_1 , \qquad l_1^+ l_1 = h_1 - \epsilon_1. \end{eqnarray}
Once we know that $l_1 l_1^+ = h_0 - \epsilon_1$ and assuming that $||\phi||^2=1$, in the real case it turns out that the normalized eigenfunctions of $h_1$ are given by 
\begin{eqnarray} \label{eq11}
\hat{\phi} = \frac{1}{\sqrt{\varepsilon - \epsilon_1}} l_1^+ \phi.
\end{eqnarray}
From now on we assume valid this expression for complex supersymmetry. 

The operator $l_1$ is useful as well to find the so called missing state, which is annihilated by $l_1$ and could be an eigenfunction of $h_1$. From the factorization $l_1^+ l_1 = h_1 - \epsilon_1$, we identify the missing state as the wave function such that $l_1 \hat{\phi}_{\epsilon_1}=0$ and $h_1 \hat{\phi}_{\epsilon_1} = \epsilon_1 \hat{\phi}_{\epsilon_1}$. By solving the first-order differential equation, it is found that 
\begin{eqnarray} \label{eq10}
\hat{\phi}_{\epsilon_1} \propto  \frac{1}{u_1}.
\end{eqnarray}

Let us stress that in complex supersymmetry $\epsilon_1 \in \mathbb{C}$ and $u_1$ is a complex function, as a result the Hermiticity of $h_1$ is not guaranteed. 

\subsection{Second-order complex supersymmetry}
We perform now a second step of the SUSY  algorithm, using an intertwining operator $l_2^+$ which intertwines $h_1$ with a Hamiltonian $h_2$ as follows: 
\begin{eqnarray}
h_2 l_2^+ = l_2^+ h_1,
\label{eq13}
\end{eqnarray}
where
\begin{eqnarray}
h_2= - \frac{d^2}{dx^2} + v_2(x), \qquad l_2^+ = - \frac{d}{dx} + \frac{\vartheta'}{\vartheta}, \label{eq14}
\end{eqnarray}
and the seed solution $\vartheta$ satisfies $-\vartheta''+v_1 \vartheta=\epsilon_2 \vartheta$. Since $\vartheta$ obeys $h_1\vartheta=\epsilon_2\vartheta$, for $\epsilon_2\neq\epsilon_1$ there must be a preimage $u_2$ such that $\vartheta=l_1^+ u_2$ fulfilling $-u_2'' + v_0 u_2 = \epsilon_2 u_2$. The potential $v_2$ thus takes the form
\begin{eqnarray} 
v_2= v_1 - 2 \frac{d^2}{dx^2}\ln \vartheta = v_0 - 2 \frac{d^2}{dx^2}(\ln u_1 \vartheta) = v_0 -2\frac{d^2}{dx^2}\ln W(u_1,u_2), 
\label{eq15}
\end{eqnarray}
where $W(f,g)=fg'-f'g$ is the Wronskian of $f$ and $g$. Moreover, combining Eqs.~\eqref{eq7} and \eqref{eq13} we can see that the second-order differential operator $l^+=l_2^+ l_1^+$ intertwines $h_0$ and $h_2$ in the way: 
\begin{eqnarray}\label{eq16}
h_2 l^+ = l^+ h_0.
\end{eqnarray}

Let us study the case where $\epsilon_2 \ne \epsilon_1$ and $v_0$ is a real potential. We would like to construct a Hermitian Hamiltonian $h_2$, thus $v_2$ has to be real. This can be accomplished by asking $W(u_1,u_2)$ to be either real or a pure imaginary function. The first case has been extensively studied elsewhere~\cite{befe12,cosc17}. The second condition can be fulfilled by asking that $u_2 = u_1^* = u^*$ and $\epsilon_2 = \epsilon_1^*=\epsilon^* \in \mathbb{C}$. 
Therefore, we can see that $W(u_1,u_2)^* = (u u^*{'}-u' u^*)^*= - W(u,u^*)$.  
Thus, by applying the first-order complex supersymmetry twice we can obtain a real potential $v_2$ and a Hermitian Hamiltonian $h_2$, even though the intermediate potential $v_1$ is complex. The second-order operator $l^+$ maps solutions of $h_0 \phi = \varepsilon \phi$ into solutions of $h_2 \widetilde{\phi} = \varepsilon \widetilde{\phi}$, where 
\begin{equation}\label{eq17}
\widetilde{\phi}= \frac{1}{\sqrt{(\varepsilon-\epsilon)(\varepsilon-\epsilon^*)}} l^+ \phi.
\end{equation}
The coefficient before $l^+ \phi$ has been included for normalization. There are now formally two missing states, one for each first-order transformation. To calculate the first one we map the missing state \eqref{eq10} of the first SUSY step using the second intertwining operator, $\widetilde{\phi}_\epsilon\propto l_2^+ (1/u_1)$. 
The second missing state is analogous to \eqref{eq10}, $\widetilde{\phi}_{\epsilon^*}\propto1/\vartheta$. Both can be expressed in terms of the seed solution $u$ as follows:
\begin{eqnarray} \label{eq18}
\widetilde{\phi}_{\epsilon} \propto \frac{u^*}{W(u,u^*)}, \quad \widetilde{\phi}_{\epsilon^*} \propto \frac{u}{W(u,u^*)}. 
\end{eqnarray}
Since we are looking for Hermitian Hamiltonians,  the missing states $\widetilde{\phi}_{\epsilon}, ~\widetilde{\phi}_{\epsilon^*}$ must not be eigenfuntions of $h_2$, i.e. they cannot be square integrable. We can accomplish this by choosing a transformation function $u$ vanishing at one end of the domain of the initial potential $v_0$. Thus, the Hamiltonian $h_2$ is isospectral to $h_0$.

\section{Complex supersymmetry in graphene: Schr\"odinger \\
equation  approach}\label{sec:Complex SUSY}

In this section we introduce step by step the notion of complex supersymmetry applied to the Dirac equation. First we note that each component of the eigenspinor of the stationary Dirac equation fulfills a Schr\"odinger equation. Then, we implement the algorithm of first-order complex supersymmetry to the Dirac Hamiltonians, where the obtained Hamiltonian is not Hermitian. It is seen that the algorithm must be iterated two more times to arrive at Hermitian Hamiltonians. We exemplify the procedure using as initial system a graphene layer in the $x-y$ plane placed in an orthogonal magnetic field of the form   $\vec{B}=\left( 0, 0,\nu k^2  \sech^2(kx)\right)$.

\subsection{Dirac equation and its supersymmetric transformation}

Let us first consider the following two-dimensional stationary Dirac equation in a magnetic field perpendicular to the $x-y$ plane: 
\begin{eqnarray} \label{Dirac 1}
H_0 \Phi = \left[ \sigma_1 \left(- i \partial_x + A_x \right)+\sigma_2 \left(- i \partial_y + A_y \right)+ m \sigma_3\right] \Phi = E \Phi,
\end{eqnarray}
where $m$ defines a mass term, $\sigma_i, ~ i=1,2,3$ are the Pauli matrices, $A_x, \ A_y$ are the components of the vector potential such that the magnetic field is $\vec{B}= B_0(x) \hat{k}= \nabla \times \vec{A}_0$, and $\Phi$ is a two-entry spinor~\cite{scro14,jk14,j15}. Suppose also that we know the solution of \eqref{Dirac 1} for a certain magnetic field, thus our goal is to generate the SUSY partner $H_1$ of $H_0$ using complex supersymmetry. To simplify the problem, we can use the so-called Landau gauge where $\vec{A}=(0,A_0(x),0)$. Since, the Hamiltonian possesses  translational symmetry along $y$-direction, it is natural to express our spinor as $\Phi=\exp (i k y) \Psi_0 = \exp (i k y) \left( \psi_0^+(x),\psi_0^-(x)\right)^T$, where $k$ is the wavenumber in $y$ direction. Thus, Eq. \eqref{Dirac 1} simplifies to: 
\begin{eqnarray} \label{eq19}
H_0 \Psi_0 = \left[ - i \sigma_1  \partial_x +\sigma_2 \left(k + A_y \right)+ m \sigma_3\right] \Psi_0 = E \Psi_0,
\end{eqnarray}
which is equivalent to the following linear system of coupled equations:
 \begin{align}
 & -i \partial_x \psi_0^- - i(k+A_0) \psi_0^- + m \psi_0^+ = E \psi_0^+, \label{SI} \\ 
 & - i \partial_x \psi_0^+ + i(k+A_0) \psi_0^+ - m \psi_0^- = E \psi_0^-. \label{S2}
 \end{align}
Solving Eq.~\eqref{S2} for $\psi_0^-$ and substituting the result in Eq.~\eqref{SI} we arrive to 
\begin{align}
&\left[ - \frac{d^2}{dx^2}  +  A_0' + (k+A_0)^2  \right] \psi_0^+ = (E^2 - m^2) \psi_0^+, \label{eq21} \\
 & \psi_0^- = \frac{i}{E+m}\left[ - \frac{d}{dx} + (k + A_0) \right] \psi_0^+.   \label{eq20} 
 \end{align}
As we can see, the upper component $\psi^+_0$ fulfills the Schr\"odinger equation: 
\begin{eqnarray}
H_0^+ \psi_0^+ = \left( - \frac{d^2}{dx^2}+V_0^+ \right) \psi_0^+ = \varepsilon \psi_0^+, \quad \varepsilon = E^2-m^2,
\label{eq22}
\end{eqnarray}
where $V_0^+= A_0' + (k + A_0)^2$. The lower component $\psi_0^-$ can be calculated using Eq.~\eqref{eq20}, once $\psi_0^+$ is known. Alternatively, we could also solve Eq.~\eqref{SI} for $\psi_0^+$ and substitute it in Eq.~\eqref{S2}, then the lower component must satisfy the Schr\"odinger equation   
\begin{eqnarray}
H_0^- \psi_0^- = \left( - \frac{d^2}{dx^2}+V_0^- \right) \psi_0^- = \varepsilon \psi_0^-, \label{otra susy} 
\end{eqnarray}
where $V_0^-= - A_0' + (k+A_0)^2 $. By comparing Eqs.~\eqref{eq22} and \eqref{otra susy}, we can see that $V_0^\pm$ are natural first-order SUSY partner potentials. If we define $k+A_0= u_0'/u_0$ and substitute it in $V_0^+=A_0' + (k + A_0)^2$, we can see that $u_0$ is in fact the seed solution of the SUSY transformation, which fulfills $-u_0''+V_0^+ u_0 = \epsilon_0 u_0$.\footnote{Typically $\epsilon_0$ is made equal to zero, since in case that $\epsilon_0\neq 0$ it is always possible to absorb it in the potential $V_0^+$ by defining $\widetilde{V}^+_0=V_0^+-\epsilon_0$.} The operator $L_0^+= - \frac{d}{dx} + \frac{u_0'}{u_0}$ intertwines $H^\pm$ as $H_0^- L_0^+ = L_0^+ H_0^+$.   

\subsection{First-order complex supersymmetry }
Let us start from equation \eqref{otra susy} by defining 
\begin{eqnarray}
H_1^+= -\frac{d^2}{dx^2} + V_1^+= H_0^-, 
\end{eqnarray}
where we are calling $V_1^+\equiv V_0^-$, $H_1^+\equiv H_0^-$. The eigenfunctions of the initial Schr\"odinger Hamiltonian $H_0^-$ are denoted as $\psi_{0,n}^-$, while the eigenfunctions to the new Hamiltonian $H_1^-$ as $\psi_{1,n}^{-}$. To build a new Dirac Hamiltonian we need $H_1^-,~\psi_{1,n}^{-},~A_1$ and $B_1$.  First we select a seed solution $u_1$ fulfilling $-u_1''+V_1^+ u_1=\epsilon_1 u_1$, or equivalently $-u_1''+V_0^- u_1= \epsilon_1 u_1$. Since we will focus on the complex SUSY algorithm, $u_1$ must be a complex function vanishing at one end of the $x$-domain of $V_0^\pm$ and $\epsilon_1 \in \mathbb{C}$. Then, the intertwining operator $L_1^+$ and the SUSY partner potential $V_1^-$ become
\begin{eqnarray}\label{1loperator}
L_1^+ = -\frac{d}{dx}+\frac{u_1'}{u_1}, \qquad V_1^-= V_1^+ - 2\frac{d^2}{dx^2} \ln u_1= V_0^- - 2\frac{d^2}{dx^2} \ln u_1, \end{eqnarray}
and $H_1^-= -\frac{d^2}{dx^2} +V_1^-$. The solutions of $H_1^- \psi_{1,n}^{-}= \varepsilon_n \psi_{1,n}^{-}$ can be found through
\begin{eqnarray}
\psi_{1,n}^{-} = \frac{i}{E_n + m} L_1^+ \psi^{-}_{0,n}, 
\end{eqnarray}
where $E^2_n = \varepsilon_n+m^2$; notice that the spectrum of $H_1^-$ is the same as $Sp(H_1^+)=Sp(H_0^-)$~\cite{lezn00,coja20}. The $y$ component of the vector potential and the magnetic field amplitude of $\vec{B}_1= B_1 \hat{k}$ are 
\begin{eqnarray} \label{A and B 1}
A_1= \frac{u_1'}{u_1}-k, \qquad B_1= \frac{d}{dx} A_1= \frac{d^2}{dx^2} \ln u_1. 
\end{eqnarray}
Finally, taking into account that $\psi_{1,n}^{+}\equiv \psi_{0,n}^{-}$ the spinor $\Psi_{1,n}= (\psi_{1,n}^{+},\psi_{1,n}^{-} )^T$ fulfills the Dirac equation
\begin{eqnarray} 
H_1 \Psi_{1,n} = \left[ - i \sigma_1  \partial_x   +\sigma_2 \left(k + A_1 \right)+ m \sigma_3\right] \Psi_{1,n} = E_n \Psi_{1,n}.
\end{eqnarray}
 It is important to remark that, in general, neither of $H_1^\pm$ are Hermitian operators because $\epsilon_1$ is complex, and thus $u_1$ is a complex function. Only when $\epsilon_1,~u_1$ become real $H_1$, could be Hermitian.

\subsection{Second-order complex supersymmetry}
Let us take now $H_1 \Psi_{1,n} = E_n \Psi_{1,n}$ as the starting problem  and repeat the complex supersymmetric algorithm. We define 
\begin{eqnarray}
H_2^+=-\frac{d^2}{dx^2} + V_2^+= H_1^-= -\frac{d^2}{dx^2}+V_0^- - 2 \frac{d^2}{dx^2} \ln u_1,
\end{eqnarray}
where the eigenfunctions $\psi_{2,n}^{+}=\psi_{1,n}^{-}$ solve the equation $H_2^+ \psi_{2,n}^{+} = \varepsilon_n \psi_{2,n}^{+}$. The next step is to choose a seed solution fulfilling $-u_2^{(1)''}+V_2^+u_2^{(1)} =\epsilon_2 u_2^{(1)}$. To obtain $u_2^{(1)}$ we use the solution $-u_2''+V_0^- u_2 = \epsilon_2 u_2$ and map it using the operator $L_1^+$ in the way 
\begin{eqnarray}
u_2^{(1)} = L_1^+ u_2 = \frac{W(u_1,u_2)}{u_1}.
\end{eqnarray}
Then 
\begin{eqnarray}
V_2^- = V_2^+ - 2 \frac{d^2}{dx^2} \ln u_2^{(1)} = V_0^- - 2 \frac{d^2}{dx^2} \ln W(u_1, u_2). \label{V2-}
\end{eqnarray}
Here we will choose $\epsilon=\epsilon_1,~u=u_1$ and fix $\epsilon_2 = \epsilon^*,~u_2= u^*$. With this selection, the potential $V_2^-$ becomes real.  
Moreover, the intertwining operator $L_2^+$ and the solutions $\psi_{2,n}^{-}$ of $H_2^- \psi_{2,n}^{-} = \varepsilon_n\psi_{2,n}^{-}$ will be given by
\begin{eqnarray} \label{psi 2 -}
L_2^+= -\frac{d}{dx}+ \frac{u_2^{(1)'}}{u_2^{(1)}}, \quad \psi_{2,n}^{-} = \frac{i}{E_n+m} L_2^+ \psi_{2,n}^{+}= -\frac{1}{(E_n+m)^2} L_2^+ L_1^+ \psi_{0,n}^{-},
\end{eqnarray}
where $E_n= \varepsilon_n+ m^2$. The vector potential $A_2$ and the magnetic field amplitude $B_2$ will be analogue expression to \eqref{A and B 1},

\begin{eqnarray} \label{A and B 2}
A_2= -k+\frac{d}{dx}\ln\frac{W(u_1, u_2)}{u_1}, \qquad B_2= \frac{d}{dx} A_2= \frac{d^2}{dx^2} \ln \frac{W(u_1, u_2)}{u_1}. 
\end{eqnarray}
Since they are still complex, it will be needed a third transformation to obtain a Hermitian Dirac Hamiltonian.   

\subsection{Third-order complex supersymmetry}

For the third SUSY step let us select $\epsilon_3$ to be a real constant. Then
\begin{eqnarray}
V_3^+ = V_2^- - \epsilon_3 = V_0^- - 2 \frac{d^2}{dx^2} \ln W(u_1, u_2) - \epsilon_3.
\end{eqnarray}
Note that $H_3^+$ is already Hermitian. The solutions of $H_3^+ \psi_{3,n}^{+}=(\varepsilon_n - \epsilon_3) \psi_{3,n}^{+}$ are the eigenfunctions $\psi_{3,n}^{+} = \psi_{2,n}^{-}$  given in Eq.~\eqref{psi 2 -}. The seed solution fulfilling $-u_3^{(2)''}+V_3^+ u_3^{(2)}=0$, or equivalent $-u_3^{(2)''}+V_2^- u_3^{(2)}=\epsilon_3 u_3^{(2)}$, can be obtained via the intertwining operators as 
$u_3^{(2)} = L_2^+ L_1^+ u_3$, where $u_3$ fulfills $-u_3''+V_0^- u_3 = \epsilon_3 u_3$. The seed solution $u_3^{(2)}$ must be a nodeless function, then $\epsilon_3 \leq \varepsilon_0$. The potential associated to $H_3^-$ reads:
\begin{eqnarray}
V_3^-=V_0^- - 2\frac{d^2}{dx^2} \ln W(u_1, u_2, u_3)-\epsilon_3. 
\end{eqnarray}
The intertwining operator $L_3^+$ and the solutions of equation $H_3^- \psi_n^{(3)-}=(\varepsilon_n - \epsilon_3) \psi_n^{(3)-}$ become 
\begin{align} \label{psi 3 -}
& L_3^+= -\frac{d}{dx}+ \frac{u_3^{(2)'}}{u_3^{(2)}}, \\  
&\psi_{3,n}^{-} = \frac{i}{\bar{E}_n+m} L_3^+ \psi_{3,n}^{+}= -\frac{i}{(\bar{E}_n+m)(E_n+m)^2} L_3^+ L_2^+ L_1^+ \psi_{0,n}^{-} 
\end{align}
where $\bar{E}_n^2=\varepsilon_n-\epsilon_3+m^2=E_n^2 - \epsilon_3$. In this case the SUSY algorithm could erase the ground state or add a new level at $\epsilon_3$. The former happens if we take $\epsilon_3=\varepsilon_0$ and the latter when the missing state $1/u_3^{(2)}$ is square integrable. The magnetic field for this step is then characterized by 
\begin{eqnarray}
A_3 =  \frac{u_3^{(2)'}}{u_3^{(2)}} - k, \quad B_3 = \frac{d}{dx} A_3 = \frac{d^2}{dx^2} \ln u_3^{(2)},
\end{eqnarray}
or, when expressed in terms of the seed solutions $u_1, u_2, u_3$, by:
\begin{eqnarray} \label{A and B 3}
A_3=-k+\frac{d}{dx}\ln\frac{W(u_1, u_2, u_3)}{W(u_1, u_2)}, \qquad B_3= \frac{d}{dx} A_3= \frac{d^2}{dx^2} \ln \frac{W(u_1, u_2, u_3)}{W(u_1, u_2)}. 
\end{eqnarray}
The spinor $\Psi_{3,n}= (\psi_{3,n}^{+},\psi_{3,n}^{-} )^T$ will solve the Dirac equation
\begin{eqnarray} \label{HD 3}
H_3 \Psi_{3,n} = \left[- i \sigma_1  \partial_x  +\sigma_2 \left(k + A_3 \right)+ m \sigma_3\right] \Psi_{3,n} = \bar{E}_n \Psi_{3,n}.
\end{eqnarray}
If in the last SUSY step we create a level in $Sp(H_3^-)$, then the spinor $\Psi_m^{(3)}\propto (0,1/u_3^{(2)} )^T$ is also a solution of the Eq.~\eqref{HD 3} associated to $\bar{E} = \pm m$. 

\subsection{Example: Graphene in a hyperbolic magnetic barrier}

\begin{figure}[t]
    \begin{subfigure}[b]{0.35\textwidth}
         \centering
         \includegraphics[width=\textwidth]{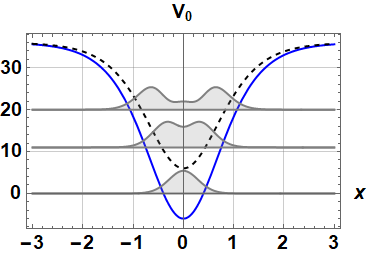}
     \end{subfigure}
         \centering
\begin{subfigure}[b]{0.35\textwidth}
         \centering  
         \includegraphics[width=\textwidth]{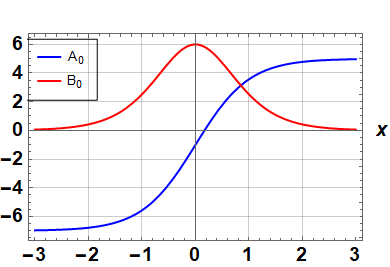}
\end{subfigure}
 \caption{Graphene in a hyperbolic magnetic barrier. (Left) Shape invariant potentials $V^-$ (blue line) and $V^+$ (dashed line). (Right) Vector potential $A_0(x)$ (blue line) and magnetic field amplitude $B_0(x)$ (red line).} \label{Fig: initial}
\end{figure}

Let us consider a vector potential $\vec{A}=(0,A_0(x),0)$ where $k+A_0=k\nu\tanh(k x)$ and $\nu > 0$. Such potential produces the magnetic field $\vec{B}=\left( 0, 0,\nu k^2  \sech^2(kx)\right)$, thus $B_0(x)=\nu k^2  \sech^2(kx)$. To solve the Dirac equation \eqref{eq19} for a massless particle, we first decouple the system of equations that the components of $\Psi_{0,n}=(\psi^+_{0,n},\psi^-_{0,n})^T$ fulfill, see Eqs.~\eqref{eq21} and \eqref{eq20}. Both components satisfy the Schr\"odinger Eqs.~\eqref{eq22} and \eqref{otra susy} with the hyperbolic P\"oschl-Teller potentials~\cite{lezn00,coja20}

\begin{equation}
   V_0^\pm= k^2\nu^2-k^2\nu(\nu\mp 1)\sech^2(k x).
\end{equation}
The general solution $\psi^-_{0}$ of Eq.~\eqref{otra susy} can be built as a superposition of the two linearly independent even and odd solutions:
\begin{eqnarray}
\psi_e & = & \cosh^{\nu+1}(k x) \, {}_2F_1\bigg(a,b,\frac{1}{2},-\sinh^2(k x)\bigg), \label{psi e} \\
    \psi_o & = & \cosh^{\nu+1}(k x)\sinh(k x) \, {}_2F_1\bigg(a+\frac{1}{2},b+\frac{1}{2},\frac{3}{2},-\sinh^2(k x)\bigg),  \label{psi o}
\end{eqnarray}
where
\begin{equation} \label{constants PT}
   a=\frac{1}{2}\left(\nu+1-\frac{\eta}{k}\right),\qquad b=\frac{1}{2}\left(\nu+1+\frac{\eta}{k}\right), \qquad \eta = \sqrt{k^2\nu^2- \varepsilon}.
\end{equation}
The well known bound state energies are given by $\varepsilon_n=k^2\nu^2-k^2(\nu-n)^2$, with $n$ being a non-negative integer such that $n < \nu$. Thus, the square integrable solutions $\psi^-_{0,n}$ are 
\begin{eqnarray} \label{psi PT}
\psi^-_{0,n} = \begin{cases} 
      \psi_e, & n \text{ even}, \\
      \psi_o, & n \text{ odd}.
   \end{cases}
\end{eqnarray}
The upper component  $\psi^+_{0,n}$ can be calculated using Eq.~\eqref{SI}. Since $m=0$, the energies of the Dirac Hamiltonian become $E_n= \pm \sqrt{k^2 \nu^2 -k^2(\nu - n)^2}$ .  
Figure \ref{Fig: initial} shows plots of the potentials $V_0^\pm$ and the probability densities $|| \Psi_{0,0}||^2,~|| \Psi_{0,1}||^2,~|| \Psi_{0,2}||^2$ (left); the corresponding vector potential and magnetic field amplitudes are shown to the right.

\begin{figure}[t]
    \begin{subfigure}[b]{0.35\textwidth}
         \centering
         \includegraphics[width=\textwidth]{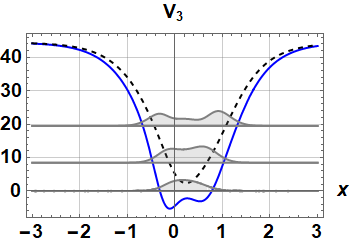}
     \end{subfigure}
         \centering
\begin{subfigure}[b]{0.35\textwidth}
         \centering  
         \includegraphics[width=\textwidth]{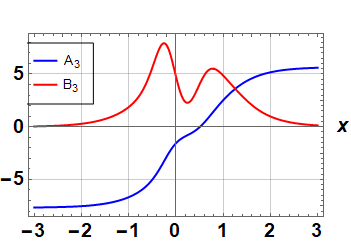}
\end{subfigure}
  \caption{Complex supersymmetry of a hyperbolic magnetic barrier. (Left) Third-order SUSY partner potentials $V^-_3$ (blue line) and $V_3^+$ (dashed line), as well as the probability densities of the first three bound states with positive energies (gray lines). (Right) Vector potential (blue line) and magnetic field amplitude (red line). The factorization energies taken are $\epsilon_1=k^2\nu^2-k^2(\nu+2i)^2$, $\epsilon_2=\epsilon_1^*$ and $\epsilon_3=k^2 \nu ^2-k^2 (\nu+2/3)^2$ with $k=1$, $\nu=6$.} \label{Fig: transformed}
\end{figure}

To apply the SUSY technique, we are going to choose as transformation functions $u_j (x,\epsilon_j)= \psi_e+ C_{j\pm} \psi_o$, satisfying the equation $H^-_0 u_j = \epsilon_j u_j$. The constants $C_{j\pm}$ read
\begin{eqnarray}
C_{j\pm}= \pm \frac{\Gamma \left(1/2\right) \Gamma\left(1-b_j \right) \Gamma \left(a_j+1/2\right)}{\Gamma \left(3/2\right) \Gamma \left(1/2-b_j \right) \Gamma(a_j)},
\end{eqnarray}
where the subscript $j=1,2,3$ depends on the SUSY step and the factorization energies $\epsilon_{1}$, $\epsilon_{2}$, $\epsilon_3$. They were carefully chosen so that $u_j(x,\epsilon_j,C_+) \rightarrow 0$ when $x \rightarrow - \infty$ and  $u_j(x,\epsilon_j,C_-) \rightarrow 0$ when $x\rightarrow \infty$. For the first SUSY step $a_1,~b_1$ take the form of Eq.~\eqref{constants PT} with $\varepsilon \rightarrow \epsilon_1\in \mathbb{C}$. Then we build the second SUSY step using the complex conjugate seed solution $u_2=u_1^*$.
The SUSY partner potentials $V_3^\pm$, obtained after the three iterations of the complex SUSY technique read: 

\begin{equation}
 \begin{split}
 V_3^+=V^-_2-\epsilon_3= k^2\nu^2-k^2\nu(\nu-1)\sech^2(kx)-2\frac{d^2}{dx^2}\ln \text{W}[{\cal F}_1(x),{\cal F}_1^*(x)]-\epsilon_3.
\end{split}
\end{equation}
\begin{equation}
 \begin{split}
 V_3^-= k^2\nu^2-k^2\nu(\nu-1)\sech^2(kx)-2\frac{d^2}{dx^2}\ln \text{W}[{\cal F}_1(x),{\cal F}_1^*(x),\widetilde{{\cal F}}_3(x)]-\epsilon_3.
\end{split}
\end{equation}
with 
\begin{equation}
{\cal F}_j(x) = {}_2F_1 [a_j,b_j,\frac{1}{2},-z^2] + C_{j\pm} z {}_2F_1[a_j+\frac{1}{2},b_j+\frac{1}{2},\frac{3}{2},-z^2],\quad j=1,2,3,
\end{equation}
and we have chosen $\widetilde{\cal F}_3={\cal F}_3(x,C_{3+})+{\cal F}_3(x,C_{3-})$. As a result, there is a normalizable ``missing state" and the energy spectrum is modified. This system  is governed by  the vector potential 
\begin{equation}
 \begin{split}
 A_3=\frac{d}{dx}\ln\frac{\text{W}[{\cal F}_1(x),{\cal F}_1^*(x),\widetilde{{\cal F}}_3(x)]}{\text{W}[{\cal F}_1(x),{\cal F}_1^*(x)]}.
\end{split}
\end{equation}
Once the vector potential is calculated, the magnetic field can be straightforwardly obtained; unfortunately its explicit expression is too long to be presented in this article. As an alternative, Figure \ref{Fig: transformed} (left) shows the transformed system arising after the three complex supersymmetry steps. The plot shows the Schr\"odinger potentials $V_3^{\pm}$ and the probability density of the first three eigenspinors of the corresponding Dirac Hamiltonian. The potential $V_3^-$ shows a clear modification as compared with $V_0^-$, meanwhile $V_3^+$ lost the parity symmetry $x \rightarrow -x$ of $V_0^+$ (compare with Figure \ref{Fig: initial}). Figure \ref{Fig: transformed} (right) shows the profiles of the vector potential and the magnetic field amplitude. 

\section{Matrix complex supersymmetry} \label{sec:Complex SUSY Matix}

Up to here, the SUSY method has been applied to the Dirac equation starting from Schr\"odinger equations. Next, we will implement an alternative SUSY technique to the Dirac equation, as it was done by Nieto,  Pecheritsin  and  Samsonov some years ago~\cite{nipesa03}, but such method will be generalized by considering complex factorization energies. Moreover, it will be shown that this matrix approach contains the algorithm exposed in Section \ref{sec:Complex SUSY}, again through three iteration steps of the formalism. Finally, a first-order matrix SUSY will be used to construct Hermitian Hamiltonians but only the zero energy modes will be found. 
\subsection{First-order matrix complex supersymmetry} 

Let us consider a one-dimensional Dirac Hamiltonian of the form:  
\begin{equation}
    H_0=-i\sigma_2\partial_y+V_0,
\label{eq26} 
\end{equation}
where $\sigma_2$ is the standard Pauli matrix, $V_0$ is an arbitrary $2\times 2$ symmetric matrix that may contain a mass term and whose entries are not necessarily real. In this section, the Dirac Hamiltonian will depend of the $y$ variable, to be consistent with~\cite{nipesa03}. However, it is possible to recover a Hamiltonian similar to Eq.~\eqref{eq19} through a unitary rotation ${\cal R}=\exp(i\pi\sigma_3/4)$ and a variable change $y\rightarrow x$ in the way
\begin{equation}
    H(x)={\cal R}H(y){\cal R}^{-1}|_{y\rightarrow x}, \qquad \Psi(x)={\cal R}\Psi(y)|_{y\rightarrow x}.
\end{equation}
The matrix SUSY approach aims to generate a new potential $V_1$ starting from $V_0$. The eigenfunctions of both Hamiltonians will be connected by the matrix \textit{intertwining} operator $\mathcal{L}^+$, except by two extra states in the new Hamiltonian $H_1$~\cite{nipesa03}. Thus, the supersymmetric Dirac Hamiltonians $H_0$ and $H_1$ fulfill the intertwining relation: 
\begin{equation}
 H_1\mathcal{L}^+=\mathcal{L}^+H_0,   
\label{eq26a}
\end{equation}
where $H_1=-i\sigma_2\partial_y+V_1$. The matrix intertwining operator is similar to the conventional operator in the Schr\"odinger approach, see Eq.~\eqref{1loperator},
\begin{equation} \label{matrixint}
    \mathcal{L}^+=-\partial_y+U_yU^{-1},
\end{equation}
with the subscript $y$ representing the derivative with respect to $y$. We assume  that $U^{-1}$ exists, thus the relation $\mathcal{L}^+U=0$ is fulfilled. Moreover, the auxiliary matrix $U$, which will be called {\it transformation matrix} or {\it matrix seed solution}, must fulfill the matrix Dirac equation 
\begin{equation}
    H_0 U=-i\sigma_2U_y+V_0U=U\Lambda,
    \label{eq27}
\end{equation}
where
\begin{equation}
U=\left(    \begin{matrix}
u_{11} & u_{12} \\
u_{21} & u_{22} 
\end{matrix}\right),\qquad
\Lambda=\left(    \begin{matrix}
\lambda_1 & 0 \\
0 & \lambda_2 
\end{matrix}\right).
\end{equation}
Here, we will focus on the case $\lambda_1, \lambda_2 \in \mathbb{C}$, which will be called \emph{matrix complex supersymmetry}. The intertwining relation (\ref{eq26a}) leads to the new potential, 
\begin{equation}
    V_1=V_0+i[U_yU^{-1},\sigma_2] =\sigma_2V_0\sigma_2+U\Lambda U^{-1}-\sigma_2U\Lambda U^{-1}\sigma_2,
\label{minter}
\end{equation}
which is also symmetric.
The intertwining operator $\mathcal{L}$ fulfilling the intertwining relation $H_0\mathcal{L}=\mathcal{L}H_1$, taking back eigenspinors of $H_1$ to eigenspinors of $H_0$, is defined as follows
\begin{equation}\label{matrixinta}
\mathcal{L}=\partial_y+(U_yU^{-1})^T,
\end{equation}
and fulfills $\mathcal{L}{(U^T)}^{-1}=0$. It is worth to note that the Hamiltonian  $H_1$ has two extra bound states, also called \emph{missing states}, corresponding to each column of the matrix $(U^T)^{-1}$.

In the Schr\"odinger approach, the intertwining operators factorize $H_0^-$ and $H_1^-$. 
There are also two factorizations in the matrix approach: the operators $\mathcal{L}^+$ and $\mathcal{L}$ factorize the Hamiltonians $H_0$ and $H_1$ as follows:  
\begin{eqnarray}
 \mathcal{L}\mathcal{L}^+ &=& (H_0-\lambda_1\mathbf{1})(H_0-\lambda_2\mathbf{1}),
\label{eq28} \\
 \mathcal{L}^+\mathcal{L} &=& (H_1-\lambda_1\mathbf{1})(H_1-\lambda_2\mathbf{1}). 
\label{eq28a}
\end{eqnarray}

This property was proven in \cite{nipesa03} for $\lambda_1,~\lambda_2$ being real numbers. For the sake of completeness, we will show that such factorizations are still valid for $\lambda_1,~\lambda_2 \in \mathbb{C}$. In order to prove Eqs.~(\ref{eq28}-\ref{eq28a}) we use the explicit expressions for $\mathcal{L}^+$ and $\mathcal{L}$ to obtain
\begin{equation}
    \mathcal{L}\mathcal{L}^+=-\partial_y^2+U_{yy}U^{-1}+\Omega(\partial_y-U_yU^{-1}),
    \label{eq59ll}
\end{equation}
where $\Omega=U_yU^{-1}-(U_yU^{-1})^T$. Making use of Eq.~(\ref{eq27}) we calculate an alternative expression for $\Omega$:
\begin{equation}
    \Omega=-i(\sigma_2V_0-(\sigma_2V_0)^T)+i(\sigma_2U\Lambda U^{-1}-(\sigma_2U\Lambda U^{-1})^T).
\end{equation}
If we consider now the symmetry of the initial potential, $V_0=V_0^T$, it is found that $\sigma_2V_0-(\sigma_2V_0)^T=0$, which leads to a simple expression for $\Omega$ in terms of the eigenvalues $\lambda_1$ and $\lambda_2$:
\begin{equation}
    \Omega=(\lambda_1+\lambda_2)i\sigma_2.
\end{equation}
Now we derive Eq.~(\ref{eq27}) with respect to $y$ and then solve for $U_{yy}U^{-1}$:
\begin{equation}
    U_{yy}U^{-1}=V_0^2-i\sigma_2V_{0y}-U\Lambda^2U^{-1}.
\end{equation}
We substitute then this expressions in the factorization formula~\eqref{eq59ll} to obtain
\begin{equation}
    \mathcal{L}\mathcal{L}^+=-\partial_y^2+V_0^2-i\sigma_2V_{0y}-(\lambda_1+\lambda_2)(-i\sigma_2\partial_y+V_0)-U\left(\Lambda^2-(\lambda_1+\lambda_2)\mathbf{1}\right)U^{-1}.
\end{equation}
Since the last term is equal to $-\lambda_1\lambda_2\mathbf{1}$, it turns out that
\begin{equation}
    \mathcal{L}\mathcal{L}^+=-\partial_y^2+V_0^2-i\sigma_2V_{0y}-(\lambda_1+\lambda_2)(-i\sigma_2\partial_y+V_0)+\lambda_1\lambda_2\mathbf{1},
\end{equation}    
which coincides with the expression we were looking for:
\begin{equation}
      \mathcal{L}\mathcal{L}^+=H_0^2-(\lambda_1+\lambda_2)H_0+\lambda_1\lambda_2.
\end{equation}
For the second factorization~(\ref{eq28a}) let us calculate $\mathcal{L}^+\mathcal{L}\mathcal{L}^+$ and use then the intertwining relation $H_1\mathcal{L}^+=\mathcal{L}^+H_0$ to obtain 
\begin{equation}
    (\mathcal{L}^+\mathcal{L})\mathcal{L}^+\Psi=(H_1-\lambda_1\mathbf{1})(H_1-\lambda_2\mathbf{1})\mathcal{L}^+\Psi.
\label{eq29}
\end{equation}
Then, the intertwining operators $\mathcal{L}^+$ and $\mathcal{L}$ supply us the simplest factorization for the partner Dirac Hamiltonians $H_0$ and $H_1$.

\subsection{Second-order matrix complex supersymmetry} \label{sec33}
The second-order matrix supersymmetry can be reached by iterating the first-order method, thus the second intertwining relation $H_2\mathcal{L}_2^+=\mathcal{L}_2^+H_1$ looks similar to its predecessor. To determine the transformation matrix ${\cal U}_2$ of the second SUSY step it is used the matrix $U_2$ that solves the equation $H_0U_2=U_2\Lambda_2$, with $\Lambda_2$ being a diagonal matrix whose elements $\widetilde{\lambda}_1,\, \widetilde{\lambda}_2$ fulfill $(\widetilde{\lambda}_1,\widetilde{\lambda}_2)\neq (\lambda_1,\lambda_2)$. The second transformation matrix that replaces the first transformation matrix ($U\rightarrow {\cal U}_2$) is built through ${\cal U}_2=\mathcal{L}^+U_2$. Therefore, the second order potential is given by the expression  
 \begin{equation}
     V_2=V_1+i\left[{\cal U}_{2y}{\cal U}_2^{-1},\sigma_2\right].
\label{eq29aa} 
\end{equation}
The second-order matrix SUSY method generates, in principle, four new eigenspinors, as compared with the original Dirac Hamiltonian $H_0$. They are given by the columns of the matrices $({\cal U}_2^T)^{-1}$ and $\mathcal{L}^+[(U^T)^{-1}]$.

\subsection{The matrix SUSY method contains the Schr\"odinger SUSY  method}
We will show that the matrix SUSY method reduces to the Schr\"odinger SUSY approach, exposed in Section \ref{sec:Complex SUSY}, for an appropriate choice of transformation matrix. Let us consider the following potential with a Dirac mass term
\begin{equation}
V_0=q_0\sigma_1+m\sigma_3. 
\end{equation}  
where $q_0=q_0(y)$. The eigenfunctions $\Psi_0 = (\psi_0^+, \psi_0^-)^T$ must satisfy the eigenvalue equation,
\begin{equation}\label{eigeninicial}
H_0 \Psi_0 = E \Psi_0,
\end{equation}
thus, the corresponding components $\psi_0^\pm$ fulfill:
\begin{equation}
(\pm\partial_y + q_0)\psi_0^\pm = (E\pm m)\psi_0^\mp \quad \Rightarrow \quad  (-\partial_y^2 + q_0^2 \mp q_0')\psi_0^\pm  
= (E^2-m^2) \psi_0^\pm .
\end{equation}
Moreover, since the transformation matrix $U$ fulfills Eq.~(\ref{eq27}) its components $u_{ij}$ must obey a similar system of equations:
\begin{eqnarray}
 u_{1j}' + q_0 \, u_{1j} & = & (\lambda_j + m) u_{2j},               \label{eq-uijk-a} \\
-u_{2j}' + q_0 \, u_{2j} & = & (\lambda_j - m) u_{1j},  \quad j=1,2. \label{eq-uijk-b}
\end{eqnarray}
These equations relate the four components of $U$ by pairs: the two equations with $j=1$ couple $u_{11}$ and $u_{21}$, while the ones with $j=2$ relate $u_{12}$ and $u_{22}$. Hence, if we supply just two components, let us say $u_{21}$ and $u_{22}$, then the other two are found through
\begin{eqnarray}
 & u_{1j} =(-u_{2j}' + q_0 \, u_{2j})/(\lambda_j - m), \quad j=1,2. 
\end{eqnarray}
Moreover, the free components $u_{21}$ and $u_{22}$ fulfill the following Schr\"odinger equations,
\begin{equation}
-u_{2j}'' + (q_0^2+q_0') \, u_{2j} = (\lambda_j^2 - m^2) u_{2j}, \quad j=1,2,
\end{equation}
while $u_{11}$ and $u_{12}$ obey the complementary Schr\"odinger equations:
\begin{equation}
-u_{1j}'' + (q_0^2-q_0') \, u_{1j} = (\lambda_j^2 - m^2) u_{1j}, \quad j=1,2.
\end{equation}

We will consider several steps of the matrix SUSY method, thus we will pick out different matrix seed solutions, which are denoted as $U_i$, associated to the matrix eigenvalues $\Lambda_i$ (we add the same superindex to the corresponding components). To apply the first SUSY step, let us choose $\Lambda_1=\text{diag}(m,-m_1)$, $m_1\in\mathbb{C}$. As we want to avoid the trivial solution we must take $u_{21}^{(1)}=0$ so that the transformation matrix $U_1$ reads
\begin{equation}
U_1=\left(    \begin{matrix}
u_{11}^{(1)} & u_{12}^{(1)} \\
0 & u_{22}^{(1)} 
\end{matrix}\right). \label{matrixseedsolution1}
\end{equation}
A straightforward calculation leads to
\begin{equation}
U_{1y}U_1^{-1}=\left(    \begin{matrix}
[\ln(u_{11}^{(1)})]' & \frac{(u_{12}^{(1)})'u_{11}^{(1)}-(u_{11}^{(1)})'u_{12}^{(1)}}{u_{11}^{(1)}u_{22}^{(1)}} \\
0 & [\ln(u_{22}^{(1)})]' 
\end{matrix}\right). \label{matrixsuperpotential}
\end{equation}
Since $u_{11}^{(1)}$, $u_{12}^{(1)}$, $u_{22}^{(1)}$ fulfill Eqs.~(\ref{eq-uijk-a}-\ref{eq-uijk-b}) with $\widetilde{\lambda}_1=m$ and $\widetilde{\lambda}_2=-m_1$, it is obtained that
\begin{equation}
U_{1y}U_1^{-1}  =\left(\begin{matrix}
-q_0 & m-m_1 \\
0 & -q_1 
\end{matrix}\right), 
\end{equation}
where $q_1\equiv-[\ln(u_{22}^{(1)})]'$. The expression (\ref{minter}) for the potential $V_1$ leads to
\begin{equation}
V_1=q_1\sigma_1+m_1\sigma_3.
\end{equation}
The components $\psi_1^\pm$ of the eigenfunction $\Psi_1 = (\psi_1^+,\psi_1^-)^T$, such that $H_1 \Psi_1 = E \Psi_1$, fulfill the new pair of Schr\"odinger equations
\begin{equation}
(-\partial_y^2 + q_1^2 \mp q_1')\psi_1^\pm  = (E^2-m_1^2) \psi_1^\pm.
\end{equation}
Since $\Psi_1 = \mathcal{L}_1^+\Psi_0$, where $\Psi_0$ satisfies Eq.~(\ref{eigeninicial}), $\psi_1^\pm$ can be expressed in terms of the corresponding components of $\Psi_0$:  
\begin{equation}
\mathcal{L}_1^+\Psi_0=(-\partial_y+U_{1y}U_1^{-1})\Psi_0 =-\left(    \begin{matrix}
 (E+m_1)\psi_0^- \\
 (\psi_0^{-})'+q_1 \psi_0^- 
\end{matrix}\right).
\end{equation}
This spinor is the analogue to the transformed eigenvector in the Schr\"odinger SUSY approach.

Let us note that the effect of the matrix SUSY transformation is to change $q_0$ by $q_1$ and $m$ by $m_1$ in $V_0$ to obtain $V_1$. This offers a clue on how to select the next matrix seed solution $U_2$ and the associated matrix eigenvalue $\Lambda_2$.
In order to implement the second SUSY step, suppose that $V_1$ is the initial potential, the transformed matrix seed solution is ${\cal U}_{2}=\mathcal{L}_1U_2$, where $U_2$ is a solution of Eq.~(\ref{eq27}) for $\Lambda_2=\text{diag}(m_1,-m_2)$, $m_2\in\mathbb{C}$. As it was discussed previously, the transformed matrix seed solution ${\cal U}_{2}$ has two free components, one of which will be chosen null, $({\cal U}_2)_{21}=0$. Since $({\cal U}_2)_{21}= W(u_{22}^{(1)},u_{21}^{(2)})/u_{22}^{(1)}$, in order to fulfill this condition we must take $u_{21}^{(2)}=u_{22}^{(1)}$. Thus, an explicit calculation leads to
\begin{equation} \label{secondseedsolution}
{\cal U}_2=\left(    \begin{matrix}
2m_1 u_{22}^{(1)} & (m_1-m_2)u_{22}^{(2)} \\
0 & W(u_{22}^{(1)},u^{(2)}_{22})/u_{22}^{(1)} 
\end{matrix}\right),
\end{equation}
which involves the two free seed solutions $u_{22}^{(1)}$ and $u_{22}^{(2)}$ used in the first and second SUSY steps respectively. Since ${\cal U}_{2}$ has the structure given in Eq.~(\ref{matrixseedsolution1}), the matrix $({\cal U}_2)_y{\cal U}_2^{-1}$ looks similar to the one of Eq.~(\ref{matrixsuperpotential}). By using now that $H_1 {\cal U}_2=
{\cal U}_2\Lambda_2$, it turns out that
\begin{equation}
{\cal U}_{2y}{\cal U}_2^{-1}  =\left(    \begin{matrix}
-q_1 & m_1-m_2 \\
0 & -q_2 
\end{matrix}\right), 
\end{equation}
where $q_2=-\partial_y\ln[{ W(u^{(1)}_{22},u^{(2)}_{22})}/{u^{(1)}_{22}}]$. A straightforward calculation leads to
the new potential $V_2$, which looks similar to $V_0$ and $V_1$:
\begin{equation}
V_2=q_2\sigma_1 + m_2\sigma_3 .
\end{equation}
The components $\psi_2^\pm$ of the eigenfunction of $H_2$, such that $H_2 \Psi_2 = E \Psi_2$, fulfill the Schr\"odinger equations
\begin{equation}
(-\partial_y^2 + q_2^2 \mp q_2')\psi_2^\pm  = (E^2-m_2^2) \psi_2^\pm.
\end{equation}
Let us note that although $u^{(2)}_{22}=(u^{(1)}_{22})^*$ and $m_2=m_1^*$, $q_2$ is still complex. Thus, it is required an extra SUSY step to obtain a Hermitian potential. Therefore, let us take as transformation matrix ${\mathfrak U}_3=\mathcal{L}_2\mathcal{L}_1 U_3$, where $U_3$ fulfills Eq.~(\ref{eq27}) with $\Lambda_3=\text{diag}(m_2,-m_3)$, $m_3\in\mathbb{R}$, and the intertwining operator reads $\mathcal{L}_3^+=-\partial_y+\mathfrak{U}_{3y}\mathfrak{U}_3^{-1}$. This time the mass parameter $m_3$ has to be real. Similarly as in the previous steps, the component $(\mathfrak{U}_3)_{21}$ must be zero. Since $(\mathfrak{U}_3)_{21}=W(u_{22}^{(1)},u_{22}^{(2)},u_{21}^{(3)})/W(u_{22}^{(1)},u_{22}^{(2)})$, we must take $u_{21}^{(3)} = u_{22}^{(2)}$ in order to fulfill this requirement. An explicit calculation leads to
\begin{eqnarray} \label{thirdseedsolution}
&& \mathfrak{U}_3=\left(    \begin{matrix}
 \frac{2m_2 W(u_{22}^{(1)},u_{22}^{(2)})}{u_{22}^{(1)}} & \frac{(m_2-m_3) W(u_{22}^{(1)},u_{22}^{(3)})}{u_{22}^{(1)}} \\
0 & \frac{W(u_{22}^{(1)},u^{(2)}_{22},u^{(3)}_{22})}{W(u_{22}^{(1)},u^{(2)}_{22})} 
\end{matrix}\right), \\
&& \mathfrak{U}_{3y}\mathfrak{U}_3^{-1}  =\left(    \begin{matrix}
-q_2 & m_2-m_3 \\
0 & -q_3 
\end{matrix}\right), 
\end{eqnarray}
where $q_3=-\partial_y\ln[{ W(u^{(1)}_{22},u^{(2)}_{22},u^{(3)}_{22})}/{W(u_{22}^{(1)},u^{(2)}_{22})]}$.    
Finally, the new Hermitian potential reads 
\begin{equation}
V_3=q_3\sigma_1-m_3\sigma_3.
\end{equation}
We can compare this potential with the one in Eq. \eqref{HD 3}, recall that the notation in this section is slightly different from Section \ref{sec:Complex SUSY}.

\subsection{Zero energy modes of graphene in electromagnetic fields via matrix complex supersymmetry}

The zero energy-modes of graphene have been studied for different settings  \cite{Ioffe19,Schulze19a,Schulze19b,Contreras20}. As final application, we will use the matrix SUSY approach with imaginary components of $\Lambda$ to find such modes. Let us start out from a Dirac Hamiltonian of the form:
\begin{eqnarray}
H_0= - i \sigma_2\partial_y + q_0 \sigma_1 + m \sigma_3,
\end{eqnarray}
where $q_0=q_0(y)$ and $m$ is a constant mass. In order to implement the SUSY transformation, let us consider the following matrices $U$ and $\Lambda$: 
\begin{equation}
U=\left(    \begin{matrix}
u_{11} & u_{12} \\
u_{21} & u_{22} 
\end{matrix}\right),\qquad
\Lambda=\left(    \begin{matrix}
\lambda_1 & 0 \\
0 & \lambda_2 
\end{matrix}\right). 
\end{equation}
Since $U$ and $\Lambda$ satisfy Eq.~\eqref{eq27}, then the components $u_{ij}$ fulfill the coupled system of Eqs.~(\ref{eq-uijk-a}-\ref{eq-uijk-b}).  
The matrix $U_yU^{-1}$ needed to perform the SUSY transformation (see Eqs.~\eqref{matrixint},\eqref{minter}) acquire the form: 
\begin{eqnarray}
U_y U^{-1} = \frac{1}{|U|}\left(    \begin{matrix}
u_{11}' & u_{12}' \\
u_{21}' & u_{22}' 
\end{matrix}\right) \left(    \begin{matrix}
u_{22} & -u_{12} \\
-u_{21} & u_{11} 
\end{matrix}\right).
\end{eqnarray}
Replacing the derivatives of $u_{ij}$ using Eqs.~(\ref{eq-uijk-a}-\ref{eq-uijk-b}), we can simplify the previous expression  
\begin{eqnarray}
U_y U^{-1} = \frac{1}{|U|} \left(f_0 \sigma_0 + f_1 \sigma_1 + f_2 \sigma_2 + f_3 \sigma_3   \right),
\end{eqnarray}
where 
\begin{eqnarray} \label{efes }
f_0 &=& \frac{1}{2}(\lambda_1 - \lambda_2) (u_{11}u_{12}+u_{21}u_{22}) ,      \nonumber \\
f_1 &=& \frac{1}{2} \left[ \left(\lambda_2- \lambda_1-2m  \right) u_{12}u_{21}+ \left(\lambda_2- \lambda_1+2m  \right) u_{11}u_{22}  \right]      \nonumber , \\
f_2 &=& \frac{1}{2 i} \left(\lambda_1+ \lambda_2 \right) \left(u_{12}u_{21}- u_{11}u_{22}   \right) = \frac{i}{2}\left(\lambda_1+ \lambda_2 \right) |U|          \nonumber ,  \\
f_3 &=& \frac{1}{2} \left\{u_{21} \left[ 2 q_0 u_{12} + (\lambda_1 - \lambda_2) u_{22}  \right] - u_{11}\left[ 2 q_0 u_{22} + (\lambda_1 - \lambda_2) u_{22}  \right]     \right\}  .   
\end{eqnarray}
The expression \eqref{minter} for the new potential $V_1$ becomes now
\begin{eqnarray} \label{v1 zero}
 V_1=V_0 + i \left[U_y U^{-1} , \sigma_2  \right] = q_0 \sigma_1 + m \sigma_3 + \frac{2 f_3 }{|U|} \sigma_1 - \frac{2 f_1 }{|U|} \sigma_3 = q_1 \sigma_1 + m_1 \sigma_3,  
\end{eqnarray}
with $q_1= q_0 + 2f_3 / |U|$ and $m_1 = m - 2 f_1 / |U|$. 

Let us restrict ourselves to the case where $\lambda_1,~\lambda_2$ are pure imaginary numbers, $m=0$ and the function $q_0$ is real. Under these conditions, and selecting $u_{21},~u_{22}$ to be real functions (we can always choose the functions $u_{21},~u_{22}$ to be real, since they are solutions of the Schr\"odinger equation $-u_{2j}''+(q_0^2+q_0') u_{2j} = \lambda_j^2 u_{2j}$ with real potential and factorization energies), the functions $q_1$ and $m_1$ become real and pure imaginary, respectively. To make sense of $H_1$ we perform the following transformation of the eigenvalue equation for the zero-energy modes,  $H_1 \Psi_{1,0} = 0$. Consider first the unitary operator $\mathcal{R}=\exp(i\pi\sigma_3/4)$, then apply the following chain of operations:  
\begin{equation}
   -i \mathcal{R} \sigma_3 H_1 \mathcal{R}^{-1}\mathcal{R}\Psi_{1,0}(y)= \widetilde{H}_1 \widetilde{\Psi}_{1,0}(y)=   \left(-i\sigma_2\partial_y+q_1 \sigma_1 -i m_1 \sigma_0 \right)\widetilde{\Psi}_{1,0}(y)=0,
\end{equation}
where $\widetilde{H}_1 =-i\mathcal{R} \sigma_3 H_1 \mathcal{R}^{-1}$, $\widetilde{V}_1= q_1 \sigma_1 - i m_1 \sigma_0$ and $\widetilde{\Psi}_{1,0}(y)= \mathcal{R}\Psi_{1,0}(y)$. Since $q_1$ and $-i m_1$ are real functions, the Hamiltonian $\widetilde{H}_1$ is Hermitian, the function $q_1$ can be associated to a magnetic field while $ -i m_1$ plays the role of an electric potential. The spinor $\widetilde{\Psi}_{1,0}= \mathcal{R}\Psi_{1,0}= \mathcal{R}\mathcal{L}^+\Psi_{0,0}$ corresponds to the zero-energy mode of this system.

\subsection{Example: Zero energy-mode of a asymtotically constant electromagnetic field} \label{zero modes}

\begin{figure}[t]
     \centering
     \begin{subfigure}[b]{0.30\textwidth}
         \centering
         \includegraphics[width=\textwidth]{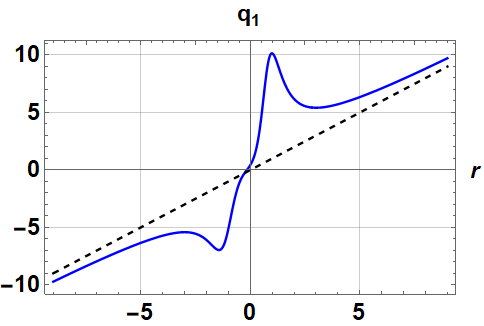}
     \end{subfigure}
     \begin{subfigure}[b]{0.30\textwidth}
         \centering
         \includegraphics[width=\textwidth]{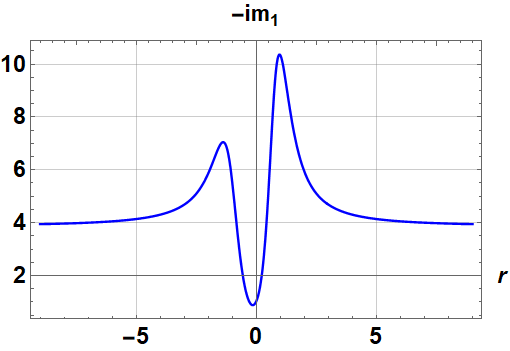}
     \end{subfigure}
          \begin{subfigure}[b]{0.30\textwidth}
         \centering
         \includegraphics[width=\textwidth]{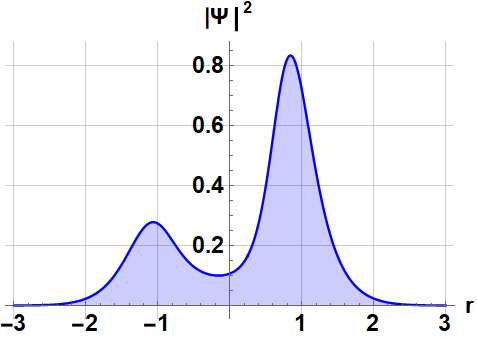}
     \end{subfigure}
          \caption{Zero-energy mode of the Hamiltonian $\widetilde{H}_1$. (Left) The magnetic terms $q_1$ in blue and $q_0=r$ in black dashed. (Center) The electric potential $-i m_1$ and (Right)  the zero energy mode $||\widetilde{\Psi}_0^{(1)}||^2$.  The parameters used in the transformation are $\lambda_1= i \sqrt{2},~\lambda_1= i \sqrt{6},~C_1=2,~C_2=-1$.  } 
        \label{graphzero}
\end{figure}

Our starting point in this example is the graphene layer in a constant magnetic field $\vec{\mathbf{A}}=(- B_0y,0,0)$. This system can be modeled through the minimal coupling rule applied to the Dirac equation, which leads to the following Hamiltonian 
\begin{equation} \label{constant field}
    H_0=-i\sigma_2\partial_y+(-i\partial_x+\frac{\alpha}{2} y)\sigma_1,
\end{equation}
where $\alpha=2eB_0/\hbar$. The eigenspinors take the form 
\begin{equation}
    \Psi_{0,n}=e^{ik_x x} \left(    \begin{matrix}
\psi^+_{0,n}(y) \\
\psi^-_{0,n-1}(y)
\end{matrix}\right),\qquad 
 \Psi_{0,0}=e^{ik_x x} \left(    \begin{matrix}
\psi^+_{0,0}(y) \\
0
\end{matrix}\right)\qquad n=1,2,\dots,
\end{equation}
with $\psi^\pm_{0,n}$ being given by
\begin{equation}
\psi^\pm_{0,n}=\frac{1}{\sqrt{2^n \pi^{1/2} n!}}e^{-\frac{1}{2}r(y)^2}H_n(r(y)),
\end{equation}
$r(y)=\sqrt{\alpha/2}(y+2k_x/\alpha)$, and $H_n(\cdot)$ are the Hermite polynomials. The Landau levels are given by $E_n=\sqrt{\alpha n}$. Using now the variable $r$, the Hamiltonian \eqref{constant field} reads 
\begin{eqnarray}
 H_0= \sqrt{\frac{\alpha}{2}} \left(-i\sigma_2 \partial_r + r \sigma_1   \right)= \sqrt{\frac{\alpha}{2}} H_{0,r}. 
\end{eqnarray}
Let us apply a first-order SUSY transformation to $H_{0,r}$ using pure imaginary constants $\lambda_1, \lambda_2$. The entries $u_{21}$ and $u_{22}$ of the transformation matrix $U$ are general solutions for the harmonic oscillator potential
\begin{equation}
u_{2j}(r)=\exp \left(-\frac{r^2}{2}\right) \left( H_{\frac{1}{2} \left(\lambda_j^2-2\right)}(y)+ C_j H_{\frac{1}{2} \left(\lambda_j^2-2\right)}(-y)\right), \quad j=1,2,
\end{equation} 
where $C_1$ and $C_2$ are real constants. The other two components of $U$ can be obtained from Eqs.~(\ref{eq-uijk-a}-\ref{eq-uijk-b}):  
\begin{equation} \label{moshinsky u11 u12}
u_{1j}=\frac{1}{\lambda_j}(-u'_{2j}+r u_{2j}).
\end{equation} 
To produce a regular transformation, we need to fix $C_1$ and $C_2$ such that det~$U \neq 0$. For this example we have used the parameters $\lambda_1= i \sqrt{2},~\lambda_2= i \sqrt{6},~C_1=2,~C_2=-1$. After $u_{11}, ~u_{12}$ are calculated through Eq.~\eqref{moshinsky u11 u12}, the direct substitution in Eqs.~\eqref{efes } and \eqref{v1 zero} provides the explicit expressions for $q_1$, $m_1$ and $\widetilde{\Psi}_{1,0}$ which, unfortunately, are lengthy. To find the zero energy mode of $H_1$ it is necessary to apply the operator $\mathcal{L}^+=-\partial_y +U_y U^{-1}$ onto the zero energy mode $\Psi_{0,0}$ of $H_0$ and then apply the unitary transformation $\mathcal{R}$. Plots of the functions $q_0, q_1, -i m_1$ and the normalized probability density $||\widetilde{\Psi}_{1,0}||^2$ are shown in Figure \ref{graphzero}.

\section{Summary}
We analyzed the behavior of charge carriers in a graphene layer under external electromagnetic fields by introducing the concept of complex supersymmetry applied to the Dirac equation using two different approaches. First, we implemented the Schr\"odinger complex SUSY approach to generate exactly solvable systems. We realized that the first SUSY step produces a non-Hermitian Dirac Hamiltonian, since the considered external magnetic field has non-trivial imaginary part. To cancel such imaginary part, it is needed to iterate the method until a third step to obtain a Hermitian Dirac Hamiltonian with a real external magnetic field $B_3$. This magnetic field shows small finite deformation with respect to the initial one, and both have the same asymptotic behavior. 
Secondly, we extended the matrix supersymmetry approach proposed by Nieto, Pecheritsin, and Samsonov \cite{nipesa03}, by including factorization energies in the complex plane.
Then, we proved that the Schr\"odinger SUSY method is a particular case of the matrix complex supersymmetry. Once again, three matrix SUSY steps are needed to recover the results of the Schr\"odinger SUSY method. As final application of the matrix SUSY method, we showed that a single matrix SUSY step plus an unitary transformation are needed to generate quasi-exactly solvable Dirac Hamiltonians for a charge carrier in graphene placed in an external electromagnetic field where its zero-energy modes are going to be calculated~\cite{cosc14,cosc17}.

\section*{Acknowledgments}
The authors acknowledge the support of Conacyt, grant FORDECYT-PRONACES/61533/2020. M. C-C. acknowledges as well the Conacyt fellowship 301117.



\begin{thebibliography}{99}
\bibitem{w47} P.R. Wallace. The band theory of graphite. {\it Phys. Rev.} {\bf 71}, 622 (1947).
\bibitem{p35} R. Peierls. Quelques propri{\'e}t{\'e}s typiques des corps solides. {\it Annales de l'institut Henri Poincar{\'e}} {\bf 5}, 177-222 (1935).

\bibitem{l37} L.D. Landau. Zur Theorie der phasenumwandlungen II. {\it Phys. Z. Sowjetunion} {\bf 11}, 26-35 (1937).

\bibitem{fv16} M.V. Fischetti and W.G. Vandenberghe. Mermin-wagner theorem, flexural modes, and degraded carrier mobility in two-dimensional crystals with broken horizontal mirror symmetry. {\it Phys. Rev. B} {\bf 93}, 155413 (2016).
\bibitem{Novoselov04a}
K. S. Novoselov, A. K. Geim, S. V. Morozov, D. Jiang, Y. Zhang, S. V. Dubonos, I. V. Grigorieva, and A. A. Firsov. Electric field effect in atomically thin carbon films. {\it Science} {\bf 306}, 666 (2004).
\bibitem{Novoselov05}
K. S. Novoselov, D. Jiang, F. Schedin, T. J. Booth, V. V. Khotkevich, S. V. Morozov,  and A. K. Geim. Two-dimensional atomic crystals. { \it Proc. Natl Acad. Sci. USA} {\bf 102}, 10451-10453 (2005). 
\bibitem{Zhang05}
Y. Zhang, Y.-W. Tan, H. L. Stormer, and P. Kim. Experimental observation of the quantum Hall effect and Berry's phase in graphene. {\it Nat.} {\bf 438}, 201-204 (2005).

\bibitem{Katsnelson06}
M. I. Katsnelson, K. S.  Novoselov, and A. K. Geim. Chiral tunnelling and the Klein paradox in graphene. {\it Nat. Phys.} {\bf 2}, 620-625 (2006).

\bibitem{be08} C.W.J. Beenakker. Colloquium: Andreev reflection and Klein tunneling in graphene. {\it Reviews of Modern Phys.} {\bf 80}, 1337 (2008).

\bibitem{CastroNeto09}
A. H. Castro Neto, F. Guinea, N. M. R. Peres, K. S. Novoselov, and A. K. Geim. The electronic properties of graphene. {\it Rev. Mod. Phys.} {\bf 81}, 109-162 (2009).

\bibitem{Lukose07}
V. Lukose, R. Shankar, and G. Baskaran. Novel Electric Field Effects on Landau Levels in Graphene. {\it Phys. Rev. Lett.} {\bf98}, 116802 (2007).

\bibitem{Stander09}
N. Stander, B. Huard, and D. Goldhaber-Gordon. Evidence for Klein Tunneling in Graphene p-n Junctions. {\it Phys. Rev. Lett.} {\bf 102}, 026807 (2009). 

\bibitem{Hartmann10}
R. R. Hartmann, N. J. Robinson, and M. E. Portnoi. Smooth electron waveguides in graphene. {\it Phys. Rev. B} {\bf 81}, 245431 (2010).

\bibitem{Kraft20}
R. Kraft, M.-H. Liu, and P. B. Selvasundaram, S.-C. Chen, R. Krupke, K. Richter,  and R. Danneau. Anomalous Cyclotron Motion in Graphene Superlattice Cavities. {\it Phys. Rev. Lett.} {\bf 125} 217701 (2020).


\bibitem{Vozmediano10}
M.A.H. Vozmediano, M. I. Katsnelson, and F. Guinea.  Gauge fields in graphene. {\it Phys. Rep.} {\bf 496}, 109-148 (2010).

\bibitem{nbo17} G. Naumis, S. Barraza-Lopez, M. Oliva-Leyva, and H. Terrones. Electronic and optical properties of strained graphene and other strained 2D materials: a review. {\it Reports on Progress Phys.} {\bf 80}, 096501 (2017).

\bibitem{Betancur18}
Y. Betancur-Ocampo. Partial positive refraction in asymmetric Veselago lenses of uniaxially strained graphene. {\it Phys. Rev. B} {\bf 98}, 205421 (2018). 

\bibitem{Contreras20a}
A. Contreras-Astorga, V. Jakubsk\'y, and A. Raya. On the propagation of Dirac fermions in graphene with strain-induced inhomogeneous Fermi velocity. {\it J. Phys. Condens. Matter} {\bf32}, 295301 (2020).

\bibitem{pdrv20}
J.C. P\'erez-Pedraza, E. D\'iaz-Bautista, A. Raya and D. Valenzuela. Critical behavior for point monopole and dipole electric impurities in uniformly and uniaxially strained graphene. {\it Phys. Rev. B.} {\bf 102}, 045131 (2020).


\bibitem{nipesa03} L.M. Nieto, A.A. Pecheritsin and B. Samsonov. Intertwining technique for the one-dimensional stationary Dirac equation. {\it Ann. Phys.}  {\bf 305}, 151-189 (2003).

\bibitem{knn09} \c{S}. Kuru, J. Negro, L.M. Nieto. Exact analitic solutions for 
a Dirac electron moving in graphene under magnetic fields. {\it J. Phys.: Condens. Matter.} {\bf 21}, 455305 (2009).

\bibitem{Jakubsky11} V. Jakubsk\'y, L.M. Nieto, and M.S. Plyushchay, Klein tunneling in carbon nanostructures: A free-particle dynamics in disguise. {\it Phys. Rev. D} {\bf 83}, 047702 (2011).

\bibitem{mf14} B. Midya and D.J. Fern\'andez C. Dirac electron in graphene under supersymmetry generated magnetic fields. {\it J. Phys. A: Math. Theor.} {\bf 47}, 285302 (2014).

\bibitem{scro14} A. Schulze-Halberg and B. Roy. Darboux partners of pseudoscalar Dirac potentials associated with exceptional orthogonal polynomials. {\it Ann. Phys.} {\bf 349}, 159-170 (2014).

\bibitem{cosc14} A. Contreras-Astorga and A. Schulze-Halberg. The confluent supersymmetry algorithm for Dirac equations with pseudoscalar potentials. {\it J. Math. Phys.} {\bf 55}, 103506 (2014).

\bibitem{Correa17}
F. Correa and V. Jakubsk\'y. Confluent Crum-Darboux transformations in Dirac Hamiltonians with PT -symmetric Bragg gratings. {\it Phys. Rev. A} {\bf 95}, 033807 (2017).


\bibitem{Ioffe19} M.V. Ioffe, D.N. Nishnianidze and E. V. Prokhvatilov. New solutions for graphene with scalar potentials by means of generalized intertwining, {\it Eur. Phys. J. Plus} {\bf 134}, 450 (2019).

\bibitem{Junker20}
G. Junker. Supersymmetric Dirac Hamiltonians in (1+1) dimensions revisited. {\it Eur. Phys. J. Plus} {\it 135} 464 (2020).

\bibitem{cefe20} M. Castillo-Celeita and D.J. Fernandez C. Dirac electron in graphene with magnetic fields arising from first-order intertwining operators. {\it J. Phys. A: Math. Theor.} {\bf 53}, 035302 (2020).

\bibitem{Contreras20} A. Contreras-Astorga, F. Correa and V. Jakubsk{\'{y}}. Super-Klein tunneling of Dirac fermions through electrostatic gratings in graphene, {\it Phys. Rev. B} {\bf 102}, 115429 (2020).

\bibitem{fgo20} D.J. Fern\'andez C., J.D. Garc\'ia, D. O-Campa.  Electron in bilayer graphene with magnetic fields leading to shape invariant potentials. {\it J. Phys. A: Math. Theor.} {\bf 53}, 435202 (2020).

\bibitem{fgo21} D.J. Fern\'andez C., J.D. Garc\'ia, D. O-Campa.  Bilayer graphene in magnetic fields generated by supersymmetry. {\it J. Phys. A: Math. Theor.} {\bf 54}, 245302 (2021).

\bibitem{mhc13} M-A. Miri, M. Heinrich, and D. Christodoulides. Supersymmetry-generated complex optical potentials with real spectra. {\it Phys. Rev. A} {\bf 87}, 043819 (2013).

\bibitem{mhe13} M-A. Miri, M. Heinrich, R. El-Ganainy, and D. Christodoulides. Supersymmetric optical structures. {\it Phys. Rev. Lett.} {\bf 110}, 233902 (2013).


\bibitem{cks95} F. Cooper, A. Khare, U. Sukhatme, Supersymmetry and quantum mechanics, {\it Phys. Rep.} {\bf 251}, 268-385 (1995). 

\bibitem{ba00} B.K. Bagchi, Supersymmetry in quantum and classical mechanics, {\it Chapman and Hall$/$CRC}, Boca Raton (2000).

\bibitem{fefe05} D.J. Fern\'andez C. and N. Fern\'andez-Garc\'ia, Higher-order supersymmetric quantum mechanics, {\it AIP Conf. Proc.} {\bf 744}, 236-273 (2005).

\bibitem{fe10} D.J. Fern\'andez C., Supersymmetric quantum mechanics, {\it AIP Conf. Proc.} {\bf 1287}, 3-36 (2010).

\bibitem{junker19}
G. Junker, Supersymmetric Methods in Quantum, Statistical and Solid State Physics. {\it IOP Publishing}, (2019).  

\bibitem{fe19}
D.J. Fern\'andez C. Trends in Supersymmetric Quantum Mechanics, in Integrability, Supersymmetry and Coherent states, \c{S}. Kuru, J. Negro, L.M. Nieto. eds., {\it CRM series in Mathematical Physics}, Springer, Cham pp. 37-68 (2019).  

\bibitem{befe12} D. Bermudez, D.J. Fern\'andez C. and N. Fern{\'a}ndez-Garc{\'\i}a. Wronskian differential formula for confluent supersymmetric quantum mechanics. {\it Phys. Lett. A} {\bf 376}, 692-696 (2012).

\bibitem{cosc17} A. Contreras-Astorga and A. Schulze-Halberg. Recursive representation of Wronskians in confluent supersymmetric quantum mechanics. {\it J. Phys. A} {\bf 50}, 105301 (2017).

\bibitem{jk14} V. Jakubsk\'y and D. Krej\v{c}i\v{r}\'ik. Qualitative analysis of trapped Dirac fermions in graphene. {\it Ann. Phys.} {\bf 349}, 268-287 (2014).

\bibitem{j15} V. Jakubsk\'y. Spectrally isomorphic Dirac systems: Graphene in an electromagnetic field. {\it Phys. Rev. D} {\bf 91}, 045039 (2015).

\bibitem{lezn00} G. L{\'e}vai and M. Znojil. Systematic search for PT-symmetric potentials with real energy spectra. {\it J. Phys. A: Math. Theor.} {\bf 33}, 7165 (2000).

\bibitem{coja20} A. Contreras-Astorga and V. Jakubsk{\'y}, Multimode Two-Dimensional PT-Symmetric Waveguides, {\it J. Phys: Conf. Ser.} {\bf 1540}, 012018 (2020).

\bibitem{Schulze19a} A. Schulze-Halberg. Higher-order Darboux transformations for the massless Dirac equation at zero energy, {\it J. Math. Phys.} {\bf 60}, 073505 (2019).

\bibitem{Schulze19b} A. Schulze-Halberg and M. Ojel. Darboux transformations for the massless Dirac equation with matrix potential: Construction of zero-energy states, {\it Eur. Phys. J. Plus} {\bf 134}, 1-12  (2019).


\end{thebibliography}
\end{document}